\documentclass[11pt,floatfix,notitlepage,byrevtex]{article}
\usepackage{graphicx,epstopdf}
\usepackage[totalwidth=500pt, totalheight=650pt]{geometry}
\usepackage{pstricks,pst-plot,pst-grad,pst-3dplot,pstricks-add,pst-node}
\usepackage{mathrsfs}   %mathscr{},...
\usepackage{amsmath}    %equation*,...
\usepackage{amssymb}    %mathfrank,hslash...
\usepackage{CJKutf8,CJKnumb}
\usepackage{hyperref}
\usepackage{CJK}
\usepackage{amsfonts}
\usepackage{caption2,graphics,placeins,verbatim}
\usepackage{diagbox}
\usepackage[numbers,sort&compress]{natbib}
\usepackage{authblk}

\allowdisplaybreaks

\title{The boundary theory of a spinor field theory on the Bruhat-Tits tree}
\author{Feng Qu\thanks{qufeng@itp.ac.cn}}
\author{Yi-hong Gao\thanks{gaoyh@itp.ac.cn}}
\affil{\small{School of Physical Sciences, University of Chinese Academy of Sciences,\\No.19A Yuquan Road, Beijing 100049, China}}
\affil{\small{CAS Key Laboratory of Theoretical Physics, Institute of Theoretical Physics, Chinese Academy of Sciences,\\Beijing 100190, China}}
\date{}

\begin{document}
\begin{CJK}{UTF8}{gbsn}
\maketitle

\begin{abstract}
For a spinor field theory on the Bruhat-Tits tree, we calculate the action and the partition function of its boundary theory by integrating out the interior of the Bruhat-Tits tree. We found that the boundary theory is very similar to a scalar field theory over $p$-adic numbers.
\end{abstract}

\section{Introduction}
\label{sec:intro}

The applications of $p$-adic analysis to string theory have provided numerous insights in the study of the AdS/CFT correspondence~\cite{Maldacena:1997re,Gubser:1998bc,Witten:1998qj}. In the early time, Freund and Olson~\cite{Freund:1987kt} considered a kind of string world sheets over $p$-adic numbers($\mathbb{Q}_p$), and gave some expressions for the $p$-adic string amplitudes. Later on, Zabrodin~\cite{Zabrodin:1988ep} found a specific realization of such kind of world sheets in term of the Bruhat-Tits tree($\textrm{T}_p$). According to Zabrodin's paper, a boundary theory, which is different from the ``CFT'' in the AdS/CFT correspondence, can be obtained by integrating out the interior of $\textrm{T}_p$. The AdS/CFT correspondence on $\textrm{T}_p$ is proposed in~\cite{Gubser:2016guj,Heydeman:2016ldy}. Some further developments are given in~\cite{Bhattacharyya:2016hbx,Gubser:2016htz,Bhattacharyya:2017aly,Gubser:2017tsi,Dutta:2017bja,Qu:2018ned,Jepsen:2018dqp,Gubser:2018cha,Hung:2018mcn,Jepsen:2018ajn,Parikh:2019ygo,Hung:2019zsk,Bentsen:2019rlr,Jepsen:2019svc}

Zabrodin only considered a massless scalar field.  Recently, the spinor field theory on $\textrm{T}_p$ has been proposed by Gubser, Jepsen and Trundy~\cite{Gubser:2018cha}. They succeed in taking the square root of the Laplacian ``$\Box$''. Let $\phi_a$ denote a field on the vertices(a vertex-field) of $\textrm{T}_p$. $\phi_a$'s on all vertices can be organized into a column vector $\phi\equiv(\phi_a,\phi_b,\phi_c,\cdots)^\textrm{T}$, where ``T'' represents the transposition. The action of $\Box$ on $\phi$ can be written as the matrix multiplication:
\begin{gather}
(\Box\phi)_a:=\sum_{b\in\partial a}(\phi_a-\phi_b)~,~(\Box)_{a,a}=p+1\textrm{~and~}(\Box)_{a,b\in\partial a}=-1~.
\end{gather}
$(\cdot)_a$ gives the entry in row $a$, and $(\cdot)_{a,b}$ gives that in row $a$ and column $b$. $b\in\partial a$ means that $b$ is one of the nearest neighboring vertices of the given vertex $a$, in other words, $b$ belongs to the boundary of $a$. Imposing a directed structure on $\textrm{T}_p$, the Laplacian has a square root $d$, which is a matrix whose row index takes value in edges and column index takes value in vertices. The result of $d$'s action on a vertex-field is a field on edges(an edge-field), and the result of $d^\textrm{T}$'s action on an edge-field is a vertex-field. Let $s(e)$ and $t(e)$ denote the starting point and the terminal point of edge $e$. The matrix $d$ satisfies
\begin{gather}
(d\phi)_e:=\phi_{t(e)}-\phi_{s(e)}~,~(d^\textrm{T}\chi)_a=\sum_{t(e)=a}\chi_e-\sum_{s(e)=a}\chi_e~,~\Box=d^\textrm{T}d~.
\end{gather}
$\chi_e$ is an edge-field. With the help of this matrix $d$ and referring to the spinor field action over real numbers, Gubser, Jepsen and Trundy propose the spinor field theory on $\textrm{T}_p$. Considering a scalar field or a spinor field on the line graph of $\textrm{T}_p$, namely L($\textrm{T}_p$), with the help of a gauge field they obtain some fermionic correlators by the AdS/CFT method.

Acknowledging Zabordin's calculation of the boundary theory and Gubser \textit{et al.}'s spinor field theory on $\textrm{T}_p$, we wonder what does the corresponding boundary theory looks like if integrating out the interior of $\textrm{T}_p$ in this spinor case. Is it a spinor-like field theory over $\mathbb{Q}_p$? Finding out its answer is the motivation of this paper. And it turns out to be a scalar-like field theory. For simplification, we only consider $\textrm{T}_p$, ignoring L($\textrm{T}_p$). All the necessary knowledge about $\mathbb{Q}_p$ and $\textrm{T}_p$ can be found in~\cite{Gubser:2016guj}. As a review of Zabordin's calculation, we consider the case of a massive scalar field, and obtain the action and the partition function of its boundary theory in the next section. Then in section \ref{sec:spinorcase}, we carry out the similar calculation in the spinor case. The last section contains a summary of our results and several unsolved problems.

In this paper we use the $(z,x)$-coordinate system for vertices on $\textrm{T}_p$, where $z=p^n~,~n\in\mathbb{Z}$ and $x\in\mathbb{Q}_p$. $\mathbb{Z}$ is the set of integers. The coordinate of a vertex $a$ writes $(z(a),x(a))$. It is the same coordinate system as the $(z_0,z)$-coordinate system in~\cite{Gubser:2016guj}. The $p$-adic norm $|\cdot|$ has the dimension of length, while the $p$-adic number itself is always dimensionless. As for the $p$-adic integration, we use the same measure $dx$ as that in~\cite{Gubser:2016guj}, which also has the dimension of length. $L$ represents the length of edges on $\textrm{T}_p$, which is a constant.

\section{The boundary theory in the scalar case}
\label{sec:scalarcase}

\begin{figure}
\setcaptionwidth{0.8\textwidth}
\centering
\includegraphics[width=0.6\textwidth]{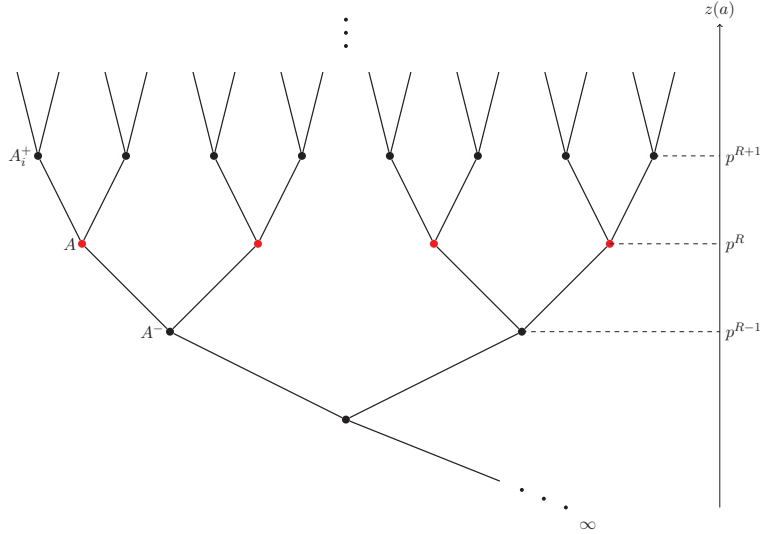}
\caption{\label{scalar}Take $\textrm{T}_{p=2}$ as an example. The red vertices compose the cut-off boundary $E_R\equiv\{A\}:=\{a|z(a)=p^R,R\in\mathbb{Z}\}$. $A^\pm$ denotes the neighboring vertices connecting to $A$ from above and below, in other words $A^\pm\in\partial A$ and $z(A^\pm)=p^{R\pm1}$. Since there are $p$ different $A^+$'s, the index $i$ is introduced to label them. The limit $R\to+\infty$ of $E_R$ gives the boundary $\mathbb{Q}_p$ where the boundary theory lives.}
\end{figure}

For a massive real-valued scalar field on $\textrm{T}_p$, two steps lead us to its boundary theory. The 1st one is that working out the partition function on a cut-off boundary $E_R$(Fig.~\ref{scalar}). The 2nd one is that taking the limit $R\to+\infty$ to obtain the partition function of the boundary theory. They are accomplished in the following two subsections.

\subsection{The partition function on $E_R$}
\label{sec:scalarEr}

The action and EOM write
\begin{gather}
S=\frac{1}{2}\sum_{\langle ab\rangle}\frac{(\phi_a-\phi_b)^2}{L^2}+\frac{1}{2}\sum_am^2\phi_a^2~,~
(\Box+L^2m^2)\phi_a=0~.
\end{gather}
$a$ or $b$ denotes the vertex and $\langle ab\rangle$ means the sum is over all the nearest neighboring vertices, in other words, over all edges.
Let $\overline{E}_R$ denote the set of all the vertices not belonging to $E_R$. Decompose $\phi$ into two, one of which is on-shell on $\overline{E}_R$, and the other one vanishes on $E_R$:
\begin{gather}
\phi_a=\Phi_a+\phi_a'~,~(\Box+L^2m^2)\Phi_a=0\textrm{~when~}a\in\overline{E}_R~,~\phi_a'=0\textrm{~when~}a\in E_R~.
\end{gather}
Due to the tree structure of $\textrm{T}_p$, $\Phi_a$ below $E_R$(``below $E_R$'' means $z(a)<p^R$) can be fixed by $\Phi_a$'s on $E_R$, namely fixed by $\Phi_A$'s. While $\Phi_a$ above $E_R$($z(a)>p^R$) can not. But we can still choose a particular configuration of $\Phi_a$'s above $E_R$ to make them also fixed by $\Phi_A$'s. Our choice is
\begin{gather}
\label{m-delta}
\textrm{~when~}z(a)>p^R~,~\Phi_a=p^{-\Delta}\Phi_{a^-}~,~L^2m^2+(1-p^{1-\Delta})(1-p^\Delta)=0~.
\end{gather}
Treating $\textrm{T}_p$ as a $p$-adic version of $\textrm{AdS}_2$, we impose the same BF bound $\Delta>1/2$ as that in~\cite{Gubser:2016guj}. It means that along the direction of $z=p^R$ to $z=\infty$, $\Phi_a$ above $E_R$ decays at the rate of $p^{-\Delta}$ per edge, which is indeed an on-shell configuration.

One comment here. Considering the Green's function $G(a,b)\propto p^{-\Delta d(a,b)}$~\cite{Gubser:2016guj}, we can write $\Phi_a$ as a superposition of $G(a,b)$'s where $b\in E_R$. $\Phi_a$ is actually a field sourced by some point sources on $E_R$. Hence the field space in this paper can be regarded as the one spanned by the Green's functions with $\Delta>1/2$.

Since all $\Phi_a$'s on $\overline{E}_R$ are fixed by $\Phi_A$'s, the measure part of the functional integral can be factorized as
\begin{gather}
\int\mathcal{D}\phi=\int_{E_R}\mathcal{D}(\Phi+\phi')\int_{\overline{E}_R}\mathcal{D}(\Phi+\phi')=\int_{E_R}\mathcal{D}\Phi\int_{\overline{E}_R}\mathcal{D}\phi'~,
\end{gather}
where $\int_{E_R}$ or $\int_{\overline{E}_R}$ means that the field only fluctuates on $E_R$ or $\overline{E}_R$.

Imposing the boundary conditions $\Phi_a,\phi_a'\to0\textrm{~when~}z(a)\to\infty$, which are consistent with the configuration (\ref{m-delta}), we can rewrite the action as
\begin{gather}
S=S_0+S'~,~S_0\equiv\frac{1}{2}\sum_{A}\Phi_A(\frac{\Box}{L^2}+m^2)\Phi_A~,~S'\equiv\frac{1}{2}\sum_a\phi_a'(\frac{\Box}{L^2}+m^2)\phi_a'~.
\end{gather}
Referring to Fig.~\ref{scalar}, the neighboring vertices of $A$ are $A^-$ and $A^+_i$'s, and that leads to $\Box\Phi_A=(p+1)\Phi_A-\Phi_{A^-}-\sum_i\Phi_{A^+_i}$. Introducing a new field $\varphi_a:=\Phi_a|z(a)|^{\Delta-1}$ and using the configuration $\Phi_{A^+_i}=p^{-\Delta}\Phi_A$, $S_0$ becomes
\begin{equation}
\label{eraction}
\begin{aligned}
S_0=&\frac{p^{1-\Delta}}{2L^2}\sum_{A}|z(A)|\varphi_A(\varphi_A-\varphi_{A^-})|z(A)|^{1-2\Delta}
\\
&+\frac{\tilde{m}^2}{2}\frac{(p^\Delta-p^{1-\Delta})L^{1-2\Delta}}{(1-p^{1-\Delta})(p^\Delta-1)}\sum_{A}|z(A)|\varphi_A^2~,
\\
\tilde{m}^2=&m^2\frac{|z(A)|^{1-2\Delta}}{L^{1-2\Delta}}~.
\end{aligned}
\end{equation}
Three comments here. First, to obtain this expression of $S_0$, we use the relation between $L,m$ and $\Delta$ in (\ref{m-delta}) following the rules which are when $R\to+\infty$, \romannumeral1)for a divergent term, make sure that there is a factor $m^2$; \romannumeral2)for a convergent term(such as $(\varphi_A-\varphi_{A^-})|z(A)|^{1-2\Delta}$ according to the next subsection), make sure that there isn't any factor $m^2$. Thus all divergent factors can be absorbed by $\tilde{m}$. These rules also apply in the spinor case. Second, when $R\to+\infty~,~\sum_{A}|z(A)|\to\int_{\mathbb{Q}_p}dx$ and $\tilde{m}^2\to\infty$, where $\tilde{m}$ can be regarded as the renormalized mass of the boundary theory. Third, we introduce $\varphi$ because it is not the $\Phi_A-\Phi_{A^-}$ term but the $\varphi_A-\varphi_{A^-}$ term that tends to a Vladimirov derivative term, which is already noticed in~\cite{Heydeman:2016ldy}.

The partition function with a source only on $E_R$ writes
\begin{equation}
\begin{aligned}
Z=&\frac{\int\mathcal{D}\phi\exp\{-S+\sum_{A}J_A\phi_A\}}
{\int\mathcal{D}\phi\exp\{-S\}}=\frac
{\int_{E_R}\mathcal{D}\varphi\exp\{-S_0+\sum_{A}|z(A)|j_A\varphi_A\}}
{\int_{E_R}\mathcal{D}\varphi\exp\{-S_0\}}~,
\\
j_A\equiv&J_A|z(A)|^{-\Delta}~.
\end{aligned}
\end{equation}
The $\int\mathcal{D}\phi'$ terms in the numerator and denominator cancel.

\subsection{The partition function of the boundary theory}
\label{sec:scalarbdy}

Suppose that $\varphi_a\to\varphi_x$ when $a\to x\in\mathbb{Q}_p$. Introduce a derivative operator in the $z$-direction $\partial_z\varphi_x:=\lim_{a\to x}(\varphi_x-\varphi_a)|z(a)|^{1-2\Delta}$. It follows that $\lim_{a\to x}(\varphi_a-\varphi_{a^-})|z(a)|^{1-2\Delta}=(p^{2\Delta-1}-1)\partial_z\varphi_x$. So when taking the limit $R\to+\infty$ or $A\to x$, we can write
\begin{gather}
S_0\to\frac{p^\Delta-p^{1-\Delta}}{2L^2}\int dx\varphi_x\partial_z\varphi_x
+\frac{\tilde{m}^2L^{1-2\Delta}(p^\Delta-p^{1-\Delta})}{2(1-p^{1-\Delta})(p^\Delta-1)}\int dx\varphi_x^2~.
\end{gather}
$\tilde{m}^2$ is divergent here. Now $\Phi_a$ becomes on-shell on the whole $\textrm{T}_p$.

We can't reconstruct the on-shell $\Phi_{A^-}$ from $\Phi_A$ in the last section, although we know the former is fixed by the latter. But when taking the limit $R\to\infty(E_R\to\textrm{the boundary~}\mathbb{Q}_p$), we can reconstruct the on-shell $\Phi_a$ using some boundary field $\Phi_x$ as
\begin{gather}\label{bulktobdy}
\Phi_a=\int dxK(a,x)\Phi_x~,~K(a,x)=\frac{p^{2\Delta}-p}{p^{2\Delta}-1}\frac{|z(a)|^\Delta}{|z(a),x(a)-x|_s^{2\Delta}}~.
\end{gather}
$K(a,x)$ is the bulk-boundary propagator in~\cite{Gubser:2016guj}, which is actually the limit of the Green's function $G(a,b)$ when $b\to x$. It is regularized by dropping a factor which tends to 0 when $b\to x$ in the cross-ratio expression of $G(a,b)$(refer to~\cite{Gubser:2016guj} for details).
$|\cdot|_s$ is the supremum norm $|x,y|_s:=sup\{|x|,|y|\}$. It follows that the reconstruction of the on-shell $\varphi_a$ writes
\begin{gather}
\varphi_a=\int dxK(a,x)|z(a)|^{\Delta-1}\Phi_x~.
\end{gather}
It can be verified that $K(a,x)|z(a)|^{\Delta-1}\to\delta(x-y)$ when $a\to y\in\mathbb{Q}_p$. So we have $\Phi_x\equiv\varphi_x$.

Using the definition of $\partial_z$ and the above reconstruction of $\varphi_a$, it can be calculated that
\begin{gather}
\partial_z\varphi_x=\frac{p^{2\Delta}-p}{p^{2\Delta}-1}\int_{y\neq x} dy\frac{\varphi_x-\varphi_y}{|x-y|^{2\Delta}}\equiv D^{2\Delta-1}\varphi_x~.
\end{gather}
$D^{2\Delta-1}$ is the $1$-dim $2\Delta-1$-th order Vladimirov derivative operator~\cite{Gubser:2016htz} up to a factor depending on $\Delta$.
Finally the action and the partition function of the boundary theory write
\begin{gather}
S_{bdy}=\frac{p^\Delta-p^{1-\Delta}}{2L^2}\int dx\varphi_xD^{2\Delta-1}\varphi_x
+\frac{\tilde{m}^2L^{1-2\Delta}(p^\Delta-p^{1-\Delta})}{2(1-p^{1-\Delta})(p^\Delta-1)}\int dx\varphi_x^2
\\
Z_{bdy}=\frac
{\int\mathcal{D}\varphi\exp\{-S_{bdy}+\int dxj_x\varphi_x\}}
{\int\mathcal{D}\varphi\exp\{-S_{bdy}\}}~,~j_x:=\lim_{A\to x}j_A~.
\end{gather}
It is actually a massive scalar field on $\mathbb{Q}_p$. The factor $p^\Delta-p^{1-\Delta}$ appears in both the kinetic and the mass term. The zero point of this factor is $\Delta=1/2$, corresponding to the BF bound.

One comment here. When we define $\partial_z$, there seems to be a free parameter $s$, which comes from $\partial_z\varphi_x:=\lim_{a\to x}(\varphi_x-\varphi_a)|z(a)|^s$. If $s\neq1-2\Delta$, we will have $\partial_z\varphi_x\times0=D^{2\Delta-1}\varphi_x$ or $\partial_z\varphi_x\times\infty=D^{2\Delta-1}\varphi_x$. If we only consider the boundary theory where both $\partial_z\varphi_x$ and $D^{2\Delta-1}\varphi_x$ are finite and have non-zero values, both cases can be ignored. So we only consider the case of $s=1-2\Delta$.

\section{The boundary theory in the spinor case}
\label{sec:spinorcase}

\begin{figure}
\setcaptionwidth{0.8\textwidth}
\centering
\includegraphics[width=0.6\textwidth]{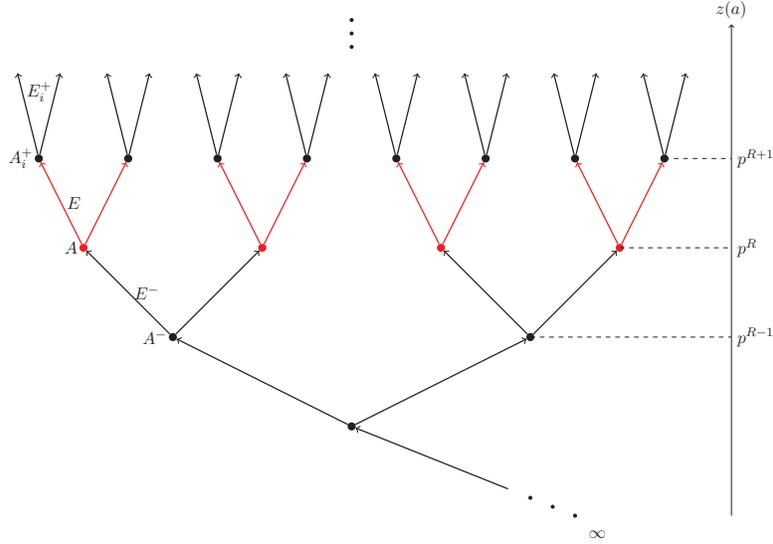}
\caption{\label{spinor}The directed $\textrm{T}_{p=2}$ tree. The direction on all edges is from bottom to top($z=0\to z=\infty$). The red vertices and edges compose the cut-off boundary $E_R\equiv\{A,E\}:=\{a,e|z(a)=p^R~,~s(e)=a\}$. $A_i^+$ and $A^-$ are the same notations as those in the scalar case. $E_i^+$ and $E^-$ are introduced to denote the edges which directly connecting to $E$ from above and below, in other words, $s(E_i^+)=t(E)$ and $t(E^-)=s(E)$.}
\end{figure}

Different from the scalar case, there is a directed structure and an edge-field besides the vertex-field on $\textrm{T}_p$. We don't need any gauge field, since it can be eliminated by redefining the vertex-field and the edge-field on a tree graph~\cite{Gubser:2018cha}. Because the difference between any two different directed structures can be absorbed by the redefined edge-field, in this paper we only consider a particular directed structure which is from bottom to top(Fig.~\ref{spinor}). We'd like to preserve both the vertex-field and the edge-field in the boundary theory, so the cut-off boundary $E_R$ must contain both vertices and edges. $E_R$ considered in this paper is shown in Fig.~\ref{spinor}. After determining $E_R$, the same steps as those in the scalar case lead us to the boundary theory.

\subsection{The partition function on $E_R$}
\label{sec:spinorEr}

The action and EOM's write
\begin{gather}
S=\frac{1}{L}\sum_e[i\chi_e^*(d\psi)_e+i\chi_e(d\psi)_e^*]+\sum_em\chi_e^*\chi_e-\sum_aM\psi_a^*\psi_a~,
\\
(id\psi)_e+Lm\chi_e=0~,~(id^\textrm{T}\chi)_a+LM\psi_a=0~.
\end{gather}
$\psi$ and $\chi$ are two Grassmann-complex-valued fields on vertices and edges. $e$ and $a$ denote the edge and the vertex.
Let $\overline{E}_R$ denotes the set of all the vertices and the edges not belonging to $E_R$. Decompose $\psi$ and $\chi$ into four fields, two of which are on-shell on $\overline{E}_R$, and the others vanish on $E_R$:
\begin{gather}
\psi_a=\Psi_a+\psi'_a~,~\chi_e=X_e+\chi'_e~,
\\
(id\Psi)_e+LmX_e=0~,~(id^\textrm{T}X)_a+LM\Psi_a=0\textrm{~when~}e,a\in\overline{E}_R~,
\\
\psi'_A=\chi'_E=0~.
\end{gather}
Similar to the scalar case, the capital fields $\Psi_a$ and $X_e$ below $E_R$ can be fixed by $\Psi_A$'s, and we can choose particular configurations for those above $E_R$, making them fixed by $X_E$'s. Be aware that the capital fields above $E_R$ can't be fixed by $\Psi_A$'s since there are off-shell $X_E$'s separating them. Considering that $\Psi_a$ above $E_R$, whose EOM writes $(d^\textrm{T} d+L^2mM)\Psi=(\Box+L^2mM)\Psi=0$, seems to be an on-shell scalar field with the mass square $mM$, we can also make it decay at the rate of $p^{-\Delta}$ per edge just as the scalar case. And $X_e$ above $E_R$ can be determined by these on-shell $\Psi_a$'s using EOM. Anyway, for the capital fields above and below $E_R$, we can write
\begin{gather}
\label{mM-delta}
\Psi_{t(E_i^+)}=p^{-\Delta}\Psi_{t(E)}~,~L^2mM+(1-p^{1-\Delta})(1-p^\Delta)=0~,
\\
i(\Psi_{t(E_i^+)}-\Psi_{t(E)})+LmX_{E_i^+}=0,\textrm{~which is the EOM on $E_i^+\in\overline{E}_R$~}~,
\\
i(X_E-\sum_iX_{E_i^+})+LM\Psi_{t(E)}=0,\textrm{~which is the EOM on $t(E)\in\overline{E}_R$~}~,
\\
i(\Psi_A-\Psi_{A^-})+LmX_{E^-}=0,\textrm{~which is the EOM on $E^-\in\overline{E}_R$~}~.
\end{gather}
Hence $\Psi_{t(E)}$ and $X_{E^-}$ can be replaced by $X_E$ and $\Psi_{A^-}$:
\begin{gather}
\label{configure}
\Psi_{t(E)}=\frac{iLm}{1-p^\Delta}X_E~,~X_{E^-}=\frac{-i}{Lm}(\Psi_A-\Psi_{A^-})~.
\end{gather}

\begin{figure}
\setcaptionwidth{0.8\textwidth}
\centering
\includegraphics[width=0.8\textwidth]{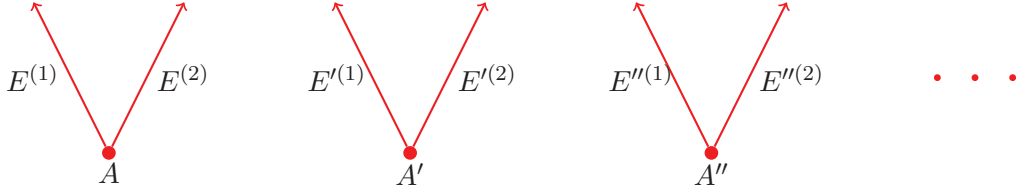}
\caption{\label{spinorER}Take $p=2$ as an example. $E_R$ is composed of red vertices and edges. There are $p$ edges connecting to the same vertex from above, which are denoted by $E^{(i)}$'s.}
\end{figure}

Imposing the boundary conditions $\Psi_a,\psi'_a,X_e,\chi'_e\to0$ when $z(a),z(t(e))\to\infty$, we can rewrite the action as
\begin{equation}
\begin{aligned}
S=&S_0+S'~,
\\
S_0=&\frac{1}{2L}\sum_{E}X_E^*[(id\Psi)_E+LmX_E]-\frac{1}{2L}\sum_{A}\Psi_A[(id^\textrm{T}X^*)_A-LM\Psi_A^*]+\textrm{c.c.}~.
\end{aligned}
\end{equation}
Here $(d\Psi)_E=\Psi_{t(E)}-\Psi_{s(E)}=\Psi_{t(E)}-\Psi_{A}$ and $(d^{\textrm{T}}X^*)_A=X_{E^-}^*-\sum_{s(E)=A}X_E^*$. Referring to Fig.~\ref{spinor}, $E^-$ is the edge below $E$. $S'$ only depends on $\psi'~,~\chi'~,~\psi'^*$, and $\chi'^*$, and ``$\textrm{c.c.}$'' represents the complex conjugate. Introduce two fields $\varphi_a:=\Psi_a|z(a)|^{\Delta-1}~,~\omega_e:=X_e|z(s(e))|^{\Delta-1}$ and one parameter $\alpha:=M/m$. Supposing that $m>0$ for taking the square root and considering the relation between $L,m,M$ and $\Delta$ in(\ref{mM-delta}) and the configurations (\ref{configure}), we can write $S_0$ as
\begin{equation}\label{adddetail}
\begin{aligned}
S_0=&\frac{-p^{1-\Delta}}{2L}c\sum_{A}|z(A)|\varphi_A^*(\varphi_A-\varphi_{A^-})|z(A)|^{1-2\Delta}
\\
&-\frac{i\tilde{m}L^{1-2\Delta}}{2}c\sum_{E}|z(A)|\omega_E^*\varphi_{A}
+\frac{i\tilde{m}L^{1-2\Delta}}{2}c\sum_{A}|z(A)|\varphi_A\sum_{s(e)=A}\omega_e^*
\\
&+\frac{\tilde{m}L^{1-2\Delta}p^\Delta}{2(p^\Delta-1)}\sum_{E}|z(A)|\omega_E^*\omega_E-\frac{\alpha\tilde{m}L^{1-2\Delta}p^\Delta}{2(p^\Delta-1)}\sum_{A}|z(A)|\varphi_A^*\varphi_A+\textrm{c.c.}~.
\end{aligned}
\end{equation}
Here we have
\begin{gather}
\tilde{m}=m\frac{|z(A)|^{1-2\Delta}}{L^{1-2\Delta}}~,~c=\sqrt{\frac{\alpha}{(1-p^{1-\Delta})(p^\Delta-1)}}~.
\end{gather}
Look into $E_R$ carefully. Referring to Fig.~\ref{spinorER}, for a general edge-field $f(E)$ we can write
\begin{equation}
\begin{aligned} &f(E^{(1)})+f(E^{(2)})+f(E'^{(1)})+f(E'^{(2)})+f(E''^{(1)})+f(E''^{(2)})+\cdots
\\
=&[f(E^{(1)})+f(E^{(2)})]+[f(E'^{(1)})+f(E'^{(2)})]+[f(E''^{(1)})+f(E''^{(2)})]+\cdots~.
\end{aligned}
\end{equation}
It indicates the relation $\sum_E=\sum_A\sum_{E^{(i)}}$. Here $\sum_{E^{(i)}}$ means the sum is over all $E^{(i)}$'s connecting to the given vertex $A$ from above. It leads to the ``$\sum_E\omega^*\varphi$'' term equals the ``$\sum_A\varphi\sum_{s(e)=A}\omega^*$'' term in (\ref{adddetail}). Hence $S_0$ also writes
\begin{equation}
\label{spinoreraction}
\begin{aligned}
&S_0=
\\
&\frac{-p^{1-\Delta}}{2L}c\sum_{A}|z(A)|\varphi_A^*(\varphi_A-\varphi_{A^-})|z(A)|^{1-2\Delta}
-i\tilde{m}L^{1-2\Delta}c\sum_{A}|z(A)|\sum_{E^{(i)}}\omega_{E^{(i)}}^*\varphi_A
\\
&+\frac{\tilde{m}L^{1-2\Delta}p^\Delta}{2(p^\Delta-1)}\sum_{A}|z(A)|\sum_{E^{(i)}}\omega_{E^{(i)}}^*\omega_{E^{(i)}}-\frac{\alpha\tilde{m}L^{1-2\Delta}p^\Delta}{2(p^\Delta-1)}\sum_{A}|z(A)|\varphi_A^*\varphi_A+\textrm{c.c.}~.
\end{aligned}
\end{equation}

Treat $\psi,\psi^*,\chi,\chi^*$ as four independent fields and introduce four corresponding sources $J^*,J,K^*,K$ on $E_R$. The measure part of the functional integral which does contribute to the finial result is the part on $E_R$(Fig.~\ref{spinorER}):
\begin{equation}
\begin{aligned}
&\int_{E_R}\mathcal{D}\psi\mathcal{D}\psi^*\mathcal{D}\chi\mathcal{D}\chi^*
=\int_A\mathcal{D}\Psi_A\mathcal{D}\Psi_A^*\prod_i\int_{E^{(i)}}\mathcal{D}X_{E^{(i)}}\mathcal{D}X_{E^{(i)}}^*
\\
=&|z(A)|^{2(p+1)(1-\Delta)}\int_A\mathcal{D}\varphi_A\mathcal{D}\varphi_A^*\prod_i\int_{E^{(i)}}\mathcal{D}\omega_{E^{(i)}}\mathcal{D}\omega_{E^{(i)}}^*~.
\end{aligned}
\end{equation}
The $\psi',\psi'^*,\chi',\chi'^*$ parts on $E_R$ don't contribute, because $\psi'_A=\chi'_E=0$. Factors $|z(A)|^{2(p+1)(1-\Delta)}$'s in the numerator and denominator of the partition function will cancel. The partition function can be written as
\begin{equation}
\begin{aligned}
&Z=
\\
&\frac{\int\mathcal{D}\psi\mathcal{D}\psi^*\mathcal{D}\chi\mathcal{D}\chi^*\exp\{-S+\sum_AJ_A^*\psi_A+\sum_EK_E^*\chi_E+\sum_A\psi_A^*J_A+\sum_E\chi_E^*K_E\}}
{\int\mathcal{D}\psi\mathcal{D}\psi^*\mathcal{D}\chi\mathcal{D}\chi^*\exp\{-S\}}
\\
&=\frac{\int_A\mathcal{D}\varphi_A\mathcal{D}\varphi_A^*(\prod_i\int_{E^{(i)}}\mathcal{D}\omega_{E^{(i)}}\mathcal{D}\omega_{E^{(i)}}^*)\exp\{-S_0+I\}}
{\int_A\mathcal{D}\varphi_A\mathcal{D}\varphi_A^*(\prod_i\int_{E^{(i)}}\mathcal{D}\omega_{E^{(i)}}\mathcal{D}\omega_{E^{(i)}}^*)\exp\{-S_0\}}~,
\\
&I\equiv\sum_A|z(A)|j_A^*\varphi_A+\sum_A|z(A)|\sum_{E^{(i)}}k_{E^{(i)}}^*\omega_{E^{(i)}}+\textrm{c.c.}
~,
\\
&j_A\equiv J_A|z(A)|^{-\Delta}~,~k_{E^{(i)}}\equiv K_{E^{(i)}}|z(A)|^{-\Delta}~.
\end{aligned}
\end{equation}

\subsection{The partition function of the boundary theory}
\label{sec:spinorbdy}

To preserve the edge-field $\omega$ in the boundary theory, we need to assign a coordinate in $\mathbb{Q}_p$ to every $E^{(i)}$. One simple choice is to set $E^{(i)}\to x$ when $A\to x\in\mathbb{Q}_p$. For example in Fig.~\ref{spinorER}, it means that $E^{(i)}\to x$ when $A\to x$, $E'^{(i)}\to x'$ when $A'\to x'$, $E''^{(i)}\to x''$ when $A''\to x''$ and so on. So in the boundary theory, there will be totally $p+1$ fields on each point, $p$ of which come from $\omega_{E^{(i)}}$'s and one of which comes from $\varphi_A$.
Considering that $\Psi_a$($\varphi_a$) is on-shell below $E_R$, which is the same as the scalar case, $\varphi_A-\varphi_{A^-}$ term can be replaced by a Vladimirov derivative term when taking the limit $R\to+\infty$ or $A\to x$. Suppose that $\omega_{E^{(i)}}\to\omega_x^{(i)}~,~j_A\to j_x~,~k_{E^{(i)}}\to k_x^{(i)}$ when $A\to x$. The boundary action and the partition function write
\begin{equation}
\begin{aligned}
S_{bdy}=&\frac{p^{1-\Delta}-p^{\Delta}}{2L}c\int dx\varphi_x^*D^{2\Delta-1}\varphi_x
-i\tilde{m}L^{1-2\Delta}c\int dx\sum_{i}(\omega_x^{(i)})^*\varphi_x
\\
&+\frac{\tilde{m}L^{1-2\Delta}p^\Delta}{2(p^\Delta-1)}\int dx\sum_{i}(\omega_x^{(i)})^*\omega_x^{(i)}-\frac{\alpha\tilde{m}L^{1-2\Delta}p^\Delta}{2(p^\Delta-1)}\int dx\varphi_x^*\varphi_x+\textrm{c.c.}~,
\\
Z_{bdy}=&\frac{\int\mathcal{D}\varphi_x\mathcal{D}\varphi_x^*(\prod_i\mathcal{D}\omega_x^{(i)}\mathcal{D}(\omega_x^{(i)})^*)\exp\{-S_{bdy}+I_{bdy}\}}
{\int\mathcal{D}\varphi_x\mathcal{D}\varphi_x^*(\prod_i\mathcal{D}\omega_x^{(i)}\mathcal{D}(\omega_x^{(i)})^*)\exp\{-S_{bdy}\}}~,
\\
I_{bdy}\equiv&\int dxj_x^*\varphi_x+\int dx\sum_{i}(k_x^{(i)})^*\omega_x^{(i)}+\textrm{c.c.}~.
\end{aligned}
\end{equation}
Similar to the scalar case, $\tilde{m}$ is also divergent. The kinetic term is the only convergent term in the action and takes the form of $\varphi^*D^{2\Delta-1}\varphi$ rather than $\omega^*D^{2\Delta-1}\varphi$. Moreover, the kinetic term of $\omega$ is missing.

Imposing the condition $\int dxD^{2\Delta-1}(\varphi_x^*\varphi_x)=0$(similar to the boundary condition of the complex-valued field theory over real numbers: $\phi_x^*\phi_x\to0\textrm{~when~}x\to\pm\infty$) and using the trick of completing the square, the functional integrals in the numerator and denominator cancel, which leads to
\begin{gather}
Z_{bdy}=\exp\{-\int dxl_x^*(\beta_1D^{2\Delta-1}+\beta_2)^{-1}l_x-\int dx\beta_3\sum_i(k_x^{(i)})^*k_x^{(i)}\}~,
\\
l_x\equiv j_x+ic(p^{-\Delta}-1)\sum_ik_x^{(i)}~,
\\
\beta_1\equiv\frac{p^\Delta-p^{1-\Delta}}{L}c~,~
\beta_2\equiv\tilde{m}L^{1-2\Delta}(p^\Delta-p^{1-\Delta})c^2~,~
\beta_3\equiv\frac{p^{-\Delta}-1}{\tilde{m}L^{1-2\Delta}}~.
\end{gather}
The factor $p^\Delta-p^{1-\Delta}$ appears in both $\beta_1$ and $\beta_2$, which is very similar to the scalar case. Further more, we can write down relations between 2-point functions as
\begin{gather}
\langle(\omega_x^{(i)})^*\varphi_y\rangle=ic(p^{-\Delta}-1)\langle\varphi_x^*\varphi_y\rangle~,~
\langle(\omega_x^{(i)})^*\omega_y^{(j)}\rangle=c^2(p^{-\Delta}-1)^2\langle\varphi_x^*\varphi_y\rangle-\delta_{ij}\beta_3~,
\end{gather}
where $\langle\varphi_x^*\varphi_y\rangle\propto((\beta_1D^{2\Delta-1}+\beta_2)^{-1})_{y,x}$, whose exact form is not considered in this paper.

\section{Summary and discussion}
\label{sec:dis}

Following~\cite{Gubser:2018cha}'s work, where a spinor field theory on $\textrm{T}_p$ is proposed, we construct the corresponding boundary theory using the method in~\cite{Zabrodin:1988ep}. We calculate the action and the partition function, and read of the relations between 2-point functions. We find that \romannumeral1)although the kinetic term of the spinor field theory on $\mathbb{T}_p$ is constructed in the form of $\chi^*d\psi$, the kinetic term of its boundary theory takes the form of $\varphi^*D^{2\Delta-1}\varphi$ rather than $\omega^*D^{2\Delta-1}\varphi$, in other words, it is a scalar-like field theory; \romannumeral2)the kinetic term of $\omega$ is missing in the action; \romannumeral3)the kinetic term is the only convergent term in the action, which is consistent with the non-renormalization theorem in~\cite{Gubser:2017vgc}; \romannumeral4)the factor $p^\Delta-p^{1-\Delta}$ appears frequently, including in the action(scalar case) and the partition function(spinor case). Its zero point corresponds to the BF bound.

There are many problems still unsolved. For example, \romannumeral1)we can't express $\varphi_{A^-}$ in (\ref{eraction}) in terms of $\varphi_A$'s. And that make it difficult to discuss the renormalization property(along the $z$-direction of $\textrm{T}_p$) of the boundary theory; \romannumeral2)the case of the line graph $\textrm{L}(\textrm{T}_p)$ equipped with a gauge field need to be considered, since only in that case the fermionic correlators can be obtained(by AdS/CFT) in both the scalar case and the spinor case. The gauge field seems to be the key element to generate fermionic correlators; \romannumeral3)we wonder what kind of theory on $\textrm{T}_p$ could lead to a ``$\omega^*D^{2\Delta-1}\varphi$''-like boundary theory. Changing the definition of $E_R$ from $\{a,e|z(a)=p^R~,~s(e)=a\}$ to $\{a,e|z(a)=p^R~,~t(e)=a\}$ doesn't help; \romannumeral4)the relation between our field space and that in~\cite{Huang:2019pgr} is not clear. Our field space is spanned by the bulk-bulk propagators $G(a,b)$'s. On the other hand, the field space in~\cite{Huang:2019pgr} is spanned by the momentum eigenfunctions $\phi^{ox\mu}_a$'s, which are also the bulk-boundary propagators. The parameter $o$ is a vertex on $\textrm{T}_p$ for regularization; $x$ is the boundary point of this propagator; $\mu$ represents the corresponding eigenvalue. $\phi^{ox\mu}_a$ is a spherically-symmetric field centered at $x$. Referring to Fig.~\ref{scalar}, take $x=\infty$ as an example and remember that a circle centered at $\infty$ is composed of all vertices with the same $z$ coordinate. We can write $\phi_a^{o\infty\mu}=\phi_b^{o\infty\mu}$ when $z(a)=z(b)$, $\phi_a^{o\infty\mu}=p^{-\mu}\phi_b^{o\infty\mu}$ when $z(a)=pz(b)$ and $\phi_a^{o\infty\mu}=1$ when $z(a)=z(o)$. We don't find out an equation relating two bulk-boundary propagators  $\phi^{ox\mu}_a$ and $K(a,x)$ in (\ref{bulktobdy}). We don't know if the bases of our field space can be written as superpositions of $\phi^{ox\mu}_a$'s;
\romannumeral5)the relation between our field space and Zabrodin's in~\cite{Zabrodin:1988ep} is not clear either. Considering that our field space is spanned by the Green's functions, which tend to 0 when approaching the boundary, the fields in our field space satisfy the boundary condition $\phi_a\to0$ when $a\to$ the boundary. On the other hand, referring to Fig.~\ref{scalar}, the counterpart of $\Phi_a$(on-shell above $E_R$) in~\cite{Zabrodin:1988ep} is a constant on each geodesic above $E_R$. So Zabrodin's fields satisfy a different boundary condition. We don't know what is the corresponding field space yet.

\section*{Acknowledgements}

We thank Miao He for useful discussions; thank the referee for important suggestions. This work is supported by the National Natural Science Foundation of China Grant No. 11947302 and 11875082.

\end{CJK}
\end{document}